\documentclass[floatfix,pra,aps,showpacs,twocolumn]{revtex4}
\usepackage{graphicx}

\begin{document}

\author{Richard J. \surname{Mathar}}
\email{mathar@strw.leidenuniv.nl}
\homepage{http://www.strw.leidenuniv.nl/~mathar}
\affiliation{
Leiden Observatory, Universiteit Leiden, Postbus 9513, 2300 RA Leiden,
The Netherlands}

\date{\today}
\title{A Table of Third and Fourth Order Feynman Diagrams
of the Interacting Fermion Green's Function}

\begin{abstract}
The Feynman diagrams of the Green's function expansion of fermions
interacting with a non-relativistic 2-body interaction are displayed
in first, second and third order of the interaction as 2, 10 and 74
diagrams, respectively. A name convention for the
diagrams is proposed and then used to tabulate the 706 diagrams
of fourth order. The Hartree-Fock approximation summons up
2, 8, 40 and 224 of them, respectively.

\end{abstract}
\maketitle

\section{Introduction}

The ground-state expectation value of the $n$-th order term in the
expansion of the Green's function
of interacting fermions
contains an integral kernel which comprises
the ground-state expectation value of the time-ordered product
of $4n+2$ field operators $\hat\psi$ and $\hat\psi^\dagger$, and 
$n$ pair interactions $U$, \cite[(8.9)]{Fetter}
\begin{eqnarray}
\langle |T[ &&
\hat\psi_{\tau}^\dagger(n)
\hat\psi_{\sigma}^\dagger(n')
U(nn')_{\tau\tau'\sigma\sigma'}
\hat\psi_{\sigma'}(n')
\hat\psi_{\tau'}(n) \nonumber\\
&& \cdots
\hat\psi_{\lambda}^\dagger(1)
\hat\psi_{\mu}^\dagger(1')
U(11')_{\lambda\lambda'\mu\mu'}
\hat\psi_{\mu'}(1')
\hat\psi_{\lambda'}(1) \nonumber\\
&& \cdot \hat\psi_{\alpha}(n+1)
\hat\psi_{\beta}^\dagger(0)]
| \rangle
.
\end{eqnarray}
The arguments of the field operators are space-time coordinates.
The two external vertices at which the test particle is created and
destroyed have been labeled $0$ and $n+1$, and the internal vertices
in pairs of primed and unprimed integers from 1 to $n$.
The Greek indices at the field operators and pair interaction
symbolize the additional dependence on spin.

We construct the Feynman diagrams \cite{FeynmanPR76_769} by contraction of the time-ordered
product with Wick's theorem \cite[\S 8]{Fetter}\cite{WickPR80},
generating all possible pairs of contractions with a computer program.
Disconnected diagrams are silently dropped during the process \cite{LuttingerPR118}\cite[\S 9]{Fetter}.

The singly connected diagrams can often be abandoned in self-consistent field
methods, like those related to Dyson's equations \cite{DysonPR75_1736},
but keeping track of them
is useful to classify all orders and an ingredient
to another hierarchical method of generation as well \cite{PelsterPSS237}.

\section{First Order and Nomenclature}

\begin{figure}[hbt]
\includegraphics[scale=0.7]{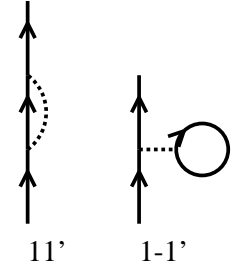}
\caption{The first order diagrams of the self-energy \cite[Fig.\ 9.7]{Fetter},
the exchange diagram at the left, the direct (or ``tadpole'') diagram
at the right.
\label{fig.1o}
}
\end{figure}

For $n=1$ we obtain two diagrams shown in Fig.\ \ref{fig.1o}.
The $2n+1$ Green's functions of the non-interacting ground state are solid
lines, the $n$ pair interactions are dashed lines.
In higher orders, the number of diagrams rises quickly, and tabulation
of the results as diagrams becomes a laborious and space-consuming process.
A tighter text version is described subsequently, and shown as a footer
to each diagram up to third order to familiarize the notation.
The fourth order diagrams in Section \ref{sec.n4} are then all
listed in the text format.

A short name will be constructed that labels each $n$-th order
diagram: let the ``backbone'' of the graph be the sequence of 2 to $2n+1$
fermion lines which leads from the external coordinate at which the
additional particle is created to the external coordinate at which it
is annihilated. This path is uniquely defined never
to divert into interaction lines;
within diagrams of this script it runs straight
upwards.
Since we will not show
disconnected diagrams, this backbone passes at least one vertex and at
most $2n$ vertices, from which interaction lines lead to fermion loops
or return to the backbone.
The vertex names are starting with $1$ at the first, and are either (i)
increased by one at the next vertex if a new interaction line starts
from there, or (ii) use the primed number of the originating vertex,
if an interaction ends that started earlier on the backbone.
(The interaction lines are not directed: the phrases ``start'' and ``end''
refer to a mere book keeping order derived from the virtual tour along all
directed fermion lines.)
The external start and end coordinates are not part of the name;
the first part of the name is the sequence of backbone vertices.
Each fermion loop is separated by a dash from the name
constructed so far; diagrams with $L$ loops have $L-1$ dashes
in the format \texttt{backbone-loop1-loop2-...-loopL}.
Within each loop a vertex is defined at which the loop is entered (see below).
The indexing is very similar to the one for the backbone: (i) either
a vertex starts a new interaction line which has not been part 
of any vertex in the backbone or any loop further to the left in the name;
then the next higher, new integer is glued to the name, or (ii) the vertex
collects an interaction line that
was started earlier in the name, in case of which the primed version of
this earlier index is attached.
Since all loops close in this kind of perturbation theory,
the ``final'' vertex within a loop equals the ``entry'' vertex: it is
not repeated in the name.
Within this scheme, the first number
after a dash---for the entry vertex---is a primed version of a number
more to the left.
As a guidance, the arrow for one Green's function within each loop
has been placed near the entry vertex in the graphs. (We tacitly assume
that all Green's functions in a loop share the same direction of rotation;
therefore one such arrow---indicative of the order of the arguments of
one Green's function---suffices.)

In addition, a hierarchical scheme of defining which of the loops is collected
first in the name (i.e., more to the left) is also needed
to maintain a one-to-one correspondence of the name to the diagram.
Loops that are reached
from the backbone by a single, direct interaction line (first order neighbors)
are listed to the left of loops that are started from intermediate loops.
Within each of the groups,
loops reached by an earlier exit
via an interaction line are placed more to the left.
If one would create a tree-type structure of the backbone and
all loops of one Feynman
Diagram, this left-right scheme would mandate that
the representation is transversed
by first handling the root (i.e., the backbone), then the first
nodes descending from the root,
then the second nodes descending from the first nodes and so on.
The vertex via which each loop is first
accessible if this ranking is introduced also defines the entry point.

The intent of these names is (i) to allow fast intuitive recovery of the
graph without loss of information in the sense that application
of the Feynman rules finds all information to build the
corresponding $2n$-fold integral
over the internal points, and (ii) one-to-one correspondence between
a name and the topology in the sense that each diagram carries only
one single name.

Side effects of the naming scheme are: (i) Each integer from 1 to $n$ shows
up once unprimed and once latter primed in the name.
(ii) There are between zero
and $n$ dashes in the name.

There are always two possible directions (clockwise, counter-clockwise)
to circle 
a loop
as a result of the semi-directed nature of the planar graphs. In low
perturbative orders (e.g., in all diagrams of Figs.\ \ref{fig.1o}--\ref{fig.3op2in1ex}),
the topology is indifferent to the choice of the direction.
In these cases the name
does not depend on this choice, and this is consistent with the fact
that then there is only one Feynman diagram to be counted and the
time-reversed (hermitian) version \emph{is} the original one.
If otherwise
the result depends on the direction to start from the loop's entry
vertex, one may (i) either flip this piece of the graph and stay with the
(for example clockwise) circulation,
as for example with the two graphs in the lower left
of Fig.\ \ref{fig.3op2in1dir},
or (ii) simply flip the arrow as done for the two pairs in the lower
right portion of Fig.\ \ref{fig.3op1in2}.

\section{Second Order}
\begin{figure}[hbt]
\includegraphics[scale=0.7]{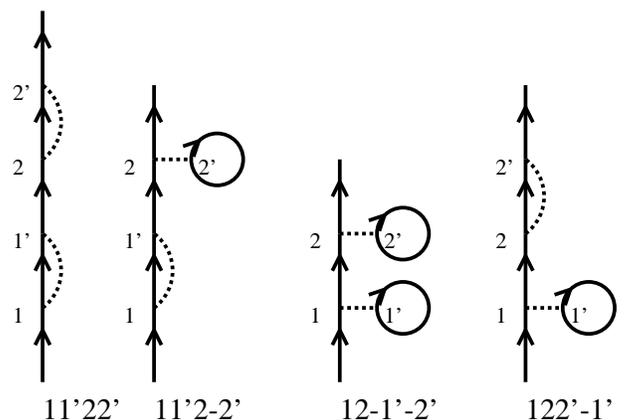}
\caption{
The $N_i(2)=4$ second order diagrams of the improper self-energy.
\label{fig.2oimp}
}
\end{figure}

\begin{figure}[hbt]
\includegraphics[scale=0.7]{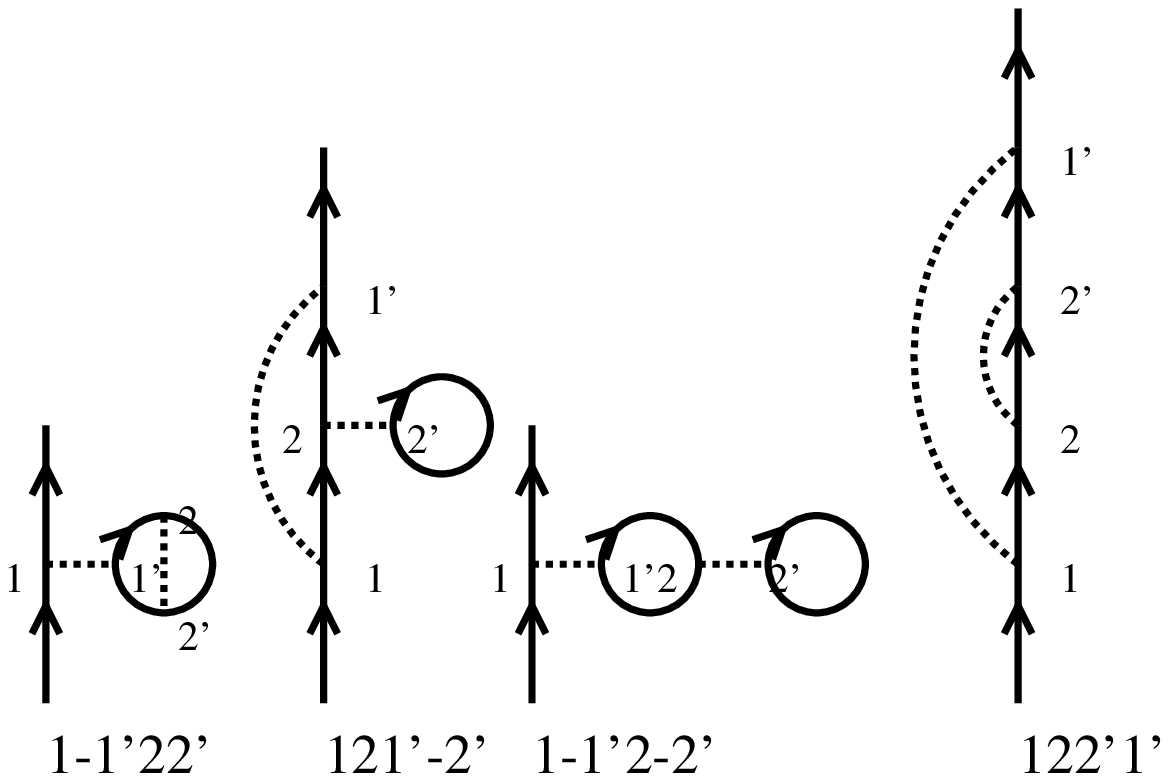}
\caption{
Second order diagrams of the proper self-energy (HF).
\label{fig.2opHF}
}
\end{figure}

\begin{figure}[hbt]
\includegraphics[scale=0.7]{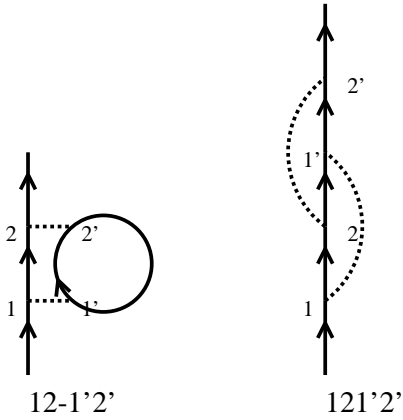}
\caption{
Second order diagrams of the proper self-energy (non-HF).
\label{fig.2opnoHF}
}
\end{figure}

The $N(2)=10$ second order diagrams, already shown in \cite[Fig. 9.8]{Fetter}
\cite[(5.8)]{BachmannPRD61} and \cite{MillsPR98},
contain $N_i(2)=4$ improper (Fig.\ \ref{fig.2oimp})
and $N_p(2)=6$ proper diagrams (Figs.\ \ref{fig.2opHF}, \ref{fig.2opnoHF}),
where
\begin{equation}
N(n)=N_i(n)+N_p(n)
\end{equation}
is the number of $n$th order diagrams, comprising
$N_i(n)$ improper and $N_p(n)$ proper diagrams.
Although formally defined in the realm of the self energy, we will
use the adjectives ``proper'' and ``improper'' for the Green's function
itself, as if being stripped of the two factors (particle lines)
that (i) create the particle at one external coordinate and annihilate
it at the first vertex, and that (ii) create the particle at the last vertex
and annihilate it at the other external coordinate.

The diagrams of Fig.\ \ref{fig.2oimp} are obtained by repeating first
order diagrams.
In any order $n\ge 1$, $2^n$ improper diagrams of this type exist because
there are $N(1)=N_p(1)=2$ first order diagrams.

\section{Third Order}
The $N(3)=74$ third order diagrams contain $N_i(3)=32$ improper
(Figs. \ref{fig.3oimp1}--\ref{fig.3oimp1plus2})
and $N_p(3)=42$ proper diagrams (Figs.\ \ref{fig.3op2in1ex}--\ref{fig.3opcmplx}).
\subsection{improper}
\label{sec.3imp}

\begin{figure}[hbt]
\includegraphics[scale=0.7]{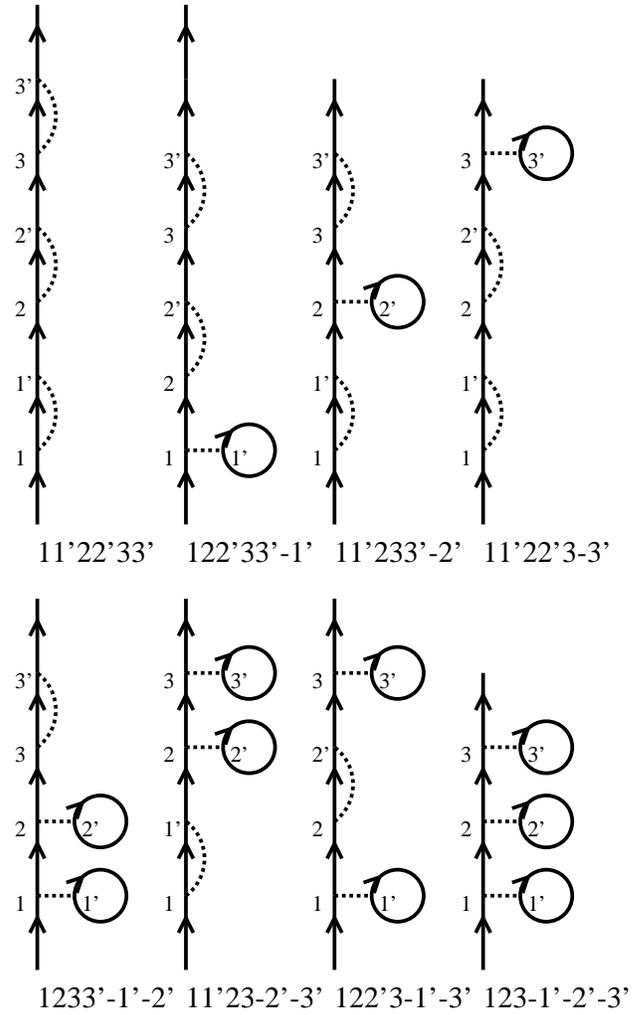}
\caption{
The 8 third order diagrams of the improper self-energy from chained first order diagrams.
\label{fig.3oimp1}
}
\end{figure}

The $n$th order improper diagrams
are generated
by concatenation of lower order diagrams, joining pairs at their
external vertices.
A classification follows
by partitioning $n$ into integer terms \cite{TouchardCJM4},
here $n=3=1+1+1=1+2=2+1$.
The diagrams of Fig.\ \ref{fig.3oimp1} are obtained by repeating three
first order diagrams according to the decomposition $1+1+1$,
producing $2\times 2\times 2=8$ diagrams.

\begin{figure}[hbt]
\includegraphics[scale=0.5]{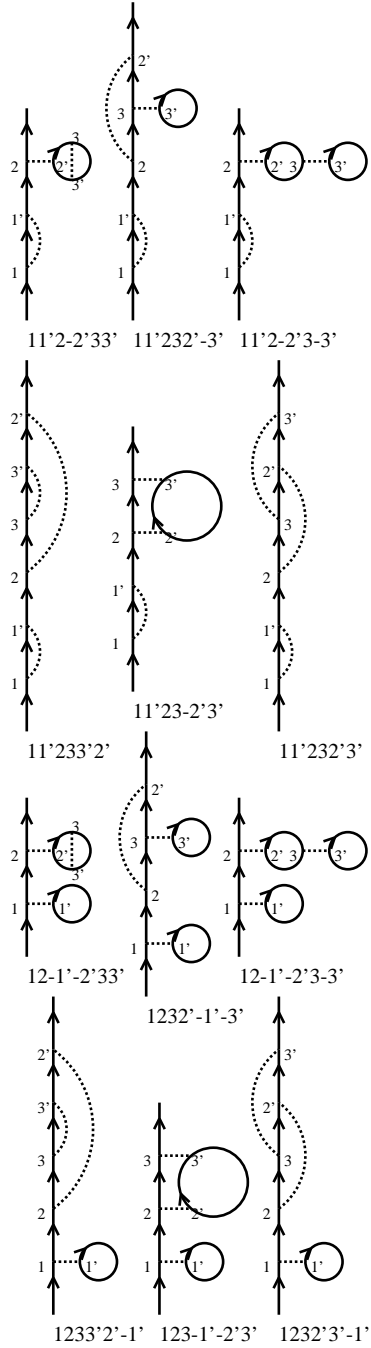}
\caption{
The 12 third order diagrams of the improper self-energy from succession of a first
order diagram and a proper second order diagram.
\label{fig.3oimp1plus2}
}
\end{figure}

\begin{figure}[hbt]
\includegraphics[scale=0.5]{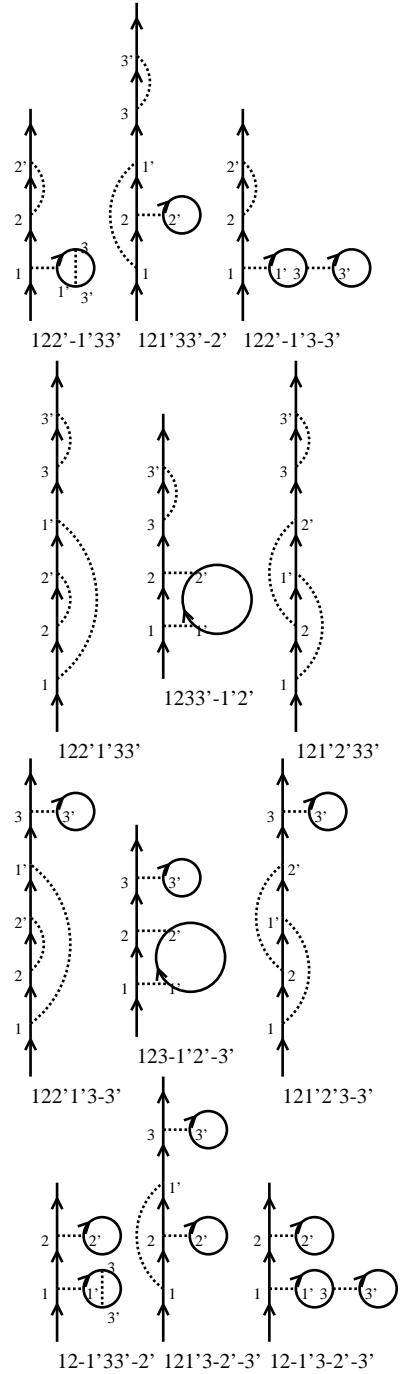}
\caption{
The 12 third order diagrams of the improper self-energy from a leading
proper second order followed by one of the two first order diagrams.
\label{fig.3oimp2plus1}
}
\end{figure}

The decomposition $1+2$ yields the remaining
$2\times 6=12$ diagrams of Fig.\ \ref{fig.3oimp1plus2} of leading first order,
and the decomposition
$2+1$ yields $6\times 2=12$ diagrams of Fig.\ \ref{fig.3oimp2plus1}.
The number $N_i(n)$ of
improper diagrams
is given by
summation over all ordered partitions weighted by the number of
diagrams of the sub-orders:
\begin{equation}
N_i(n)=\sum_{l=1}^{n-1} N_p(l)N(n-l).
\label{eq.Ni}
\end{equation}
There are $N_p(l)$ ways to choose a ``leading'' proper diagram that
contains what we have labeled vertex 1, and each can be followed in
$N(n-l)$ ways by some diagram of order $n-l$.

\subsection{proper}
\label{sec.3op}
\begin{figure}[hbt]
\includegraphics[scale=0.6]{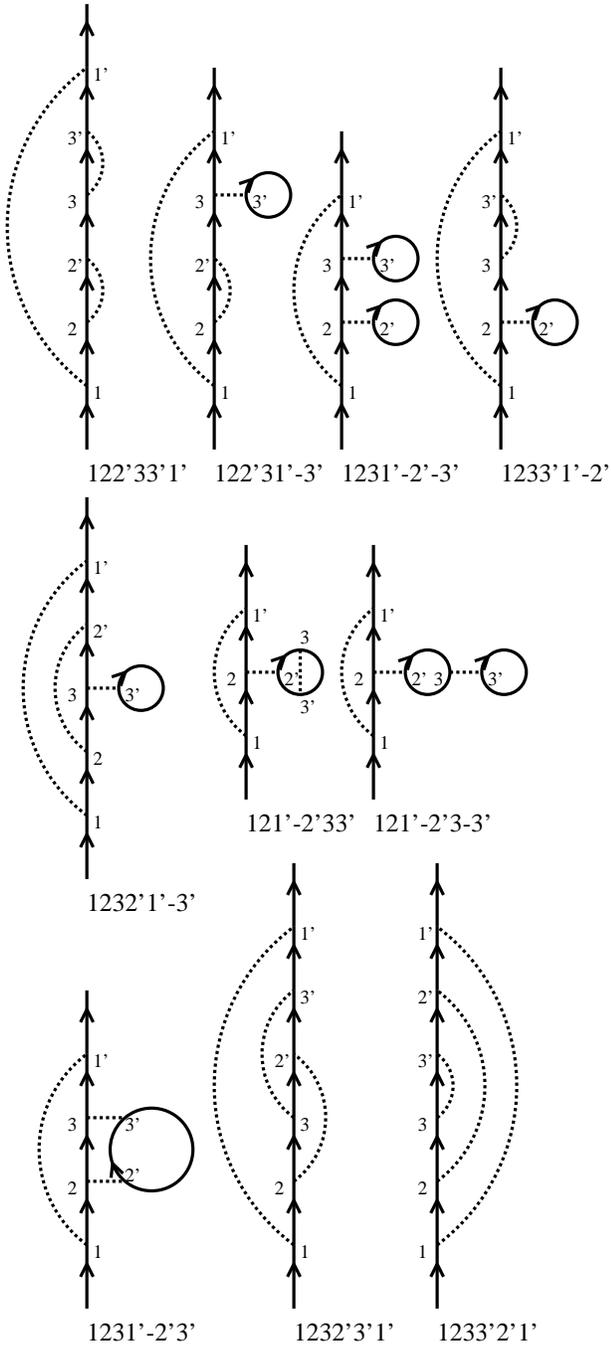}
\caption{
The third order diagrams of the proper self-energy from insertion of the 10
second order diagrams into the first order 11' ``exchange'' diagram.
Two diagrams,
1231'-2'3' and 1232'3'1', are shown as
F12 and F14 in \cite[Fig 4]{BlundellPRA42}.
\label{fig.3op2in1ex}
}
\end{figure}

\begin{figure}[hbt]
\includegraphics[scale=0.6]{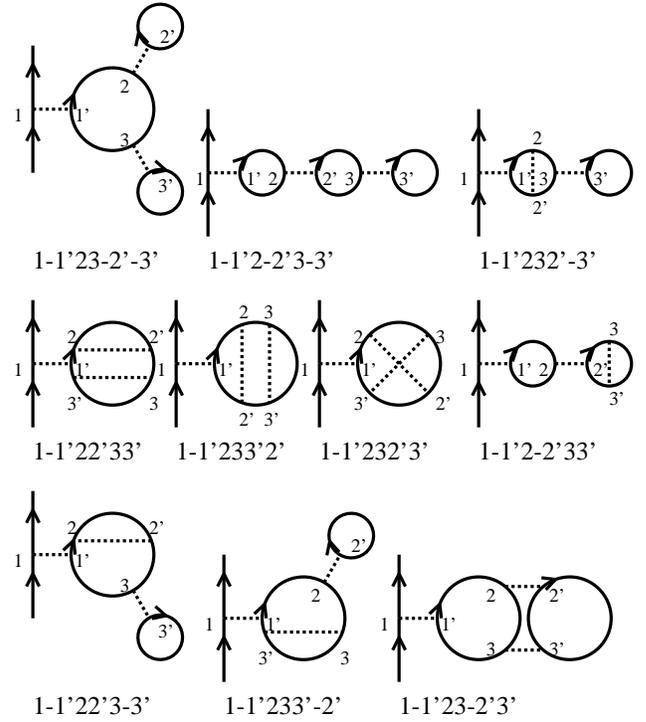}
\caption{
The third order diagrams of the proper self-energy from insertion of the 10
second order diagrams into the first order 1-1' ``direct'' diagram.
Two of these,
1-1'232'3' and 1-1'23-2'3', are shown as
F11 and F13 in \cite[Fig 4]{BlundellPRA42}.
\label{fig.3op2in1dir}
}
\end{figure}

The third order diagrams in Fig.\ \ref{fig.3op2in1ex} are created by
blending the second order diagrams into the first order diagram 11' which
becomes the skeleton.
The names are derived from the second order ``intruder'' diagrams by incrementing
all numbers in the intruder diagram by one and enclosing the backbone part of
the name in a pair of 1 and 1'.
The third order diagrams in Fig.\ \ref{fig.3op2in1dir} are created by
blending the second order diagrams into the first order tadpole diagram:
In accordance with the rules described above,
the names are derived from the second order intruder diagrams by incrementing
all numbers in the intruder diagram by one, then placing 1-1' in front.
(Two examples: 1-1'22'33' from 11'22', and 1-1'22'3-3'
from 11'2-2'.)

\begin{figure}[hbt]
\includegraphics[scale=0.55]{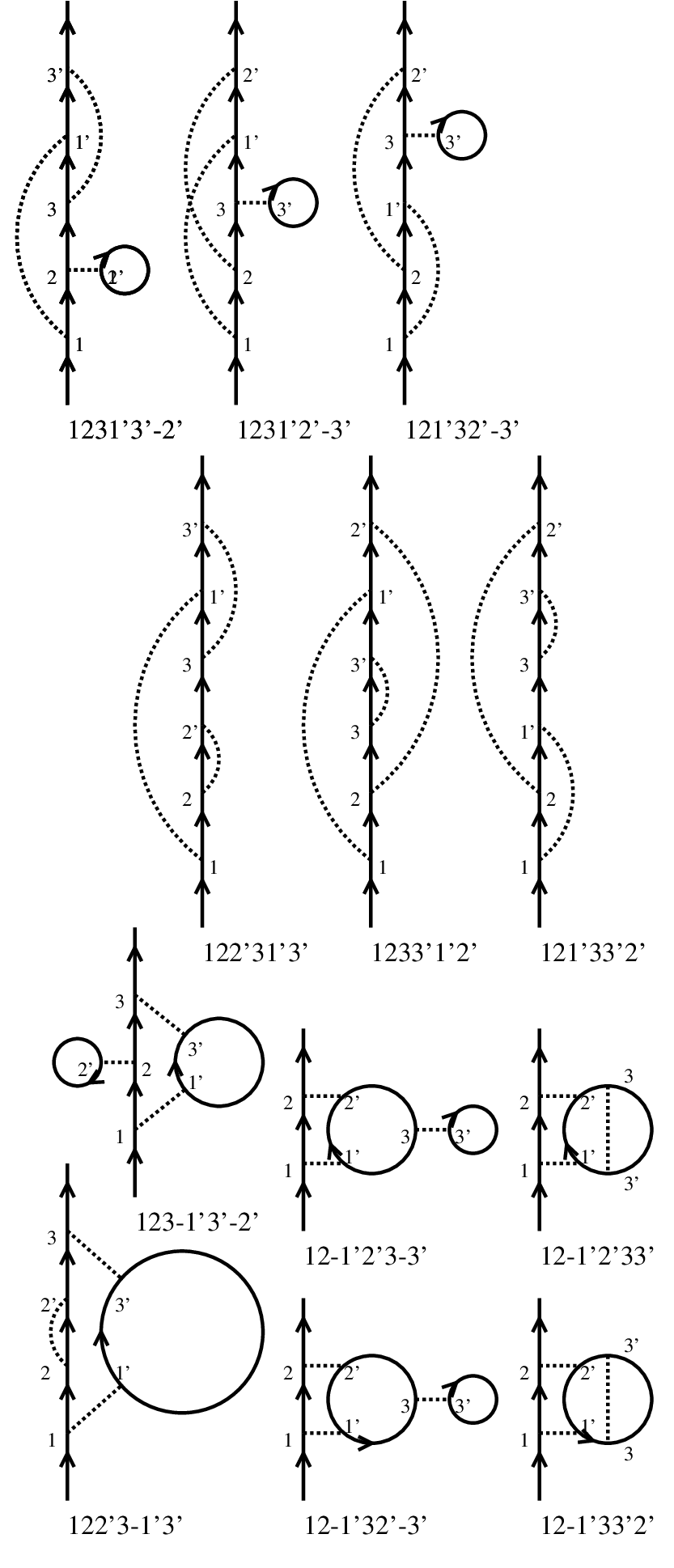}
\caption{
Third order diagrams of the proper self-energy from insertion of a first
order diagram into a second order fermion line.
\label{fig.3op1in2}
}
\end{figure}

More doubly-connected
third order diagrams follow from insertion of any of the two first 
order diagrams into one of the three internal fermion lines of the
proper second order diagrams: Fig.\ \ref{fig.3op1in2} shows those that have not
yet been part of Fig.\ \ref{fig.3op2in1dir}.

\begin{figure}[hbt]
\includegraphics[scale=0.6]{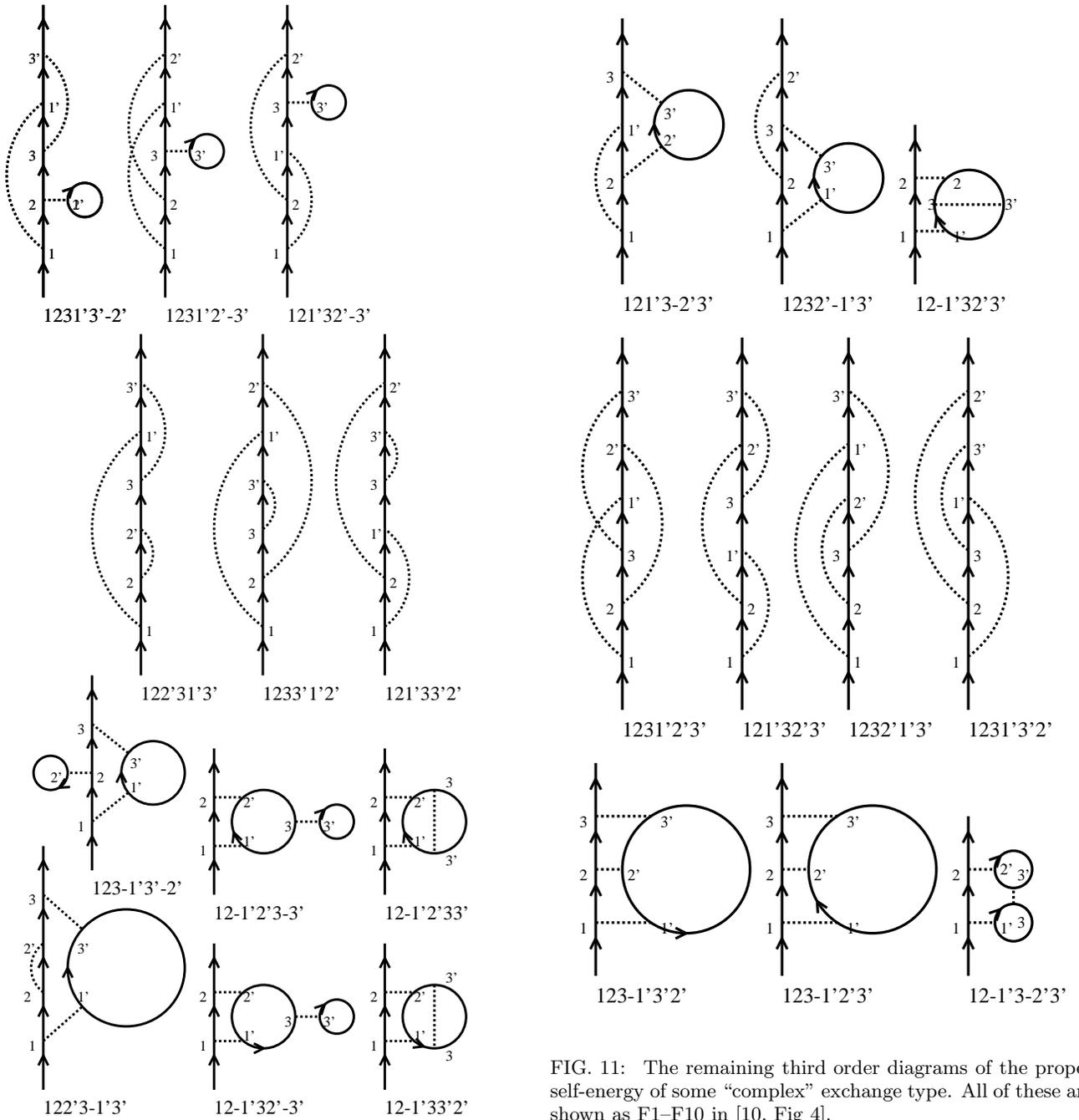}
\caption{
The remaining third order diagrams of the proper self-energy of some
``complex'' exchange type.
All of these are shown as F1--F10  in \cite[Fig 4]{BlundellPRA42}.
\label{fig.3opcmplx}
}
\end{figure}

The concluding set of third order proper diagrams in Fig.\ \ref{fig.3opcmplx}
contains diagrams that cannnot be decomposed into a skeleton
with insertions.

\clearpage
\section{Fourth order}
\label{sec.n4}
The $N(4)=706$ diagrams of the fourth order perturbation contain
$N_i(4)=2\times 74+6\times 10+2\times 42=292$
improper diagrams of leading order $l=1,2$ or $3$ according
to (\ref{eq.Ni}), and $N_p(4)=414$ proper diagrams.

\subsection{improper}
Guided by the partitioning of the order 4 into $4=1+\ldots=2+\ldots=3+1$,
the improper diagrams are either (i) a first order diagram followed
by a proper or improper third order diagram, or (ii) a proper second order
diagram followed by a  proper or improper second order diagram, or (iii)
a proper third order diagram followed by a first order diagram.
This classification leads to the following subsections.

\subsubsection{Leading First Order}
\label{sec.4o1l}
This set contains $N_p(1)N(3)=2\times 74$ diagrams since we have 2 first order
diagrams and 74 third order diagrams. If the first order diagram is the
exchange diagram, the name pattern is 11'\ldots and is listed
in Table \ref{tab.4oiL1ex}; if the first order
diagram is the direct diagram, the name pattern is 1\ldots-1'-\ldots
or 1\ldots-1' and is listed in Table \ref{tab.4oiL1dir}.

\begin{ruledtabular}
\begin{table}[hbt]
\caption{
74 improper fourth order diagrams created by starting with 11' and
attaching one of the 32 improper third order diagrams
(first eight lines) or 42 proper third order diagrams (final eleven lines)
\label{tab.4oiL1ex}
}
\begin{tabular}{llll}
11'233'2'4-4'
&
11'233'2'44'
&
11'2344'3'-2'
&
11'22'344'3'
\\
11'23-2'44'-3'
&
11'233'-2'44'
&
11'23-2'-3'44'
&
11'22'3-3'44'
\\
11'22'34-3'-4'
&
11'22'344'-3'
&
11'233'4-2'-4'
&
11'22'33'4-4'
\\
11'22'33'44'
&
11'22'343'4'
&
11'22'34-3'4'
&
11'22'3-3'4-4'
\\
11'233'44'-2'
&
11'22'343'-4'
&
11'234-2'-3'-4'
&
11'2344'-2'-3'
\\
11'2343'4'-2'
&
11'234-2'-3'4'
&
11'23-2'-3'4-4'
&
11'2343'-2'-4'
\\
11'232'3'4-4'
&
11'232'3'44'
&
11'234-2'3'-4'
&
11'2344'-2'3'
\\
11'23-2'4-3'-4'
&
11'233'-2'4-4'
&
11'232'4-3'-4'
&
11'232'44'-3'
\\
\hline
11'2344'3'2'
&
11'2-2'344'3'
&
11'232'-3'44'
&
11'233'42'-4'
\\
11'2-2'33'4-4'
&
11'233'44'2'
&
11'23-2'3'44'
&
11'233'4-2'4'
\\
11'2-2'33'44'
&
11'2344'2'3'
&
11'232'44'3'
&
11'2344'2'-3'
\\
11'233'42'4'
&
11'23-2'44'3'
&
11'2-2'344'-3'
&
11'234-2'3'4'
\\
11'2342'3'4'
&
11'2342'3'-4'
&
11'23-2'3'4-4'
&
11'23-2'43'4'
\\
11'2342'4'3'
&
11'2342'4'-3'
&
11'23-2'4-3'4'
&
11'2343'-2'4'
\\
11'234-2'4'-3'
&
11'23-2'43'-4'
&
11'2343'2'4'
&
11'234-2'4'3'
\\
11'232'43'-4'
&
11'232'4-3'4'
&
11'232'43'4'
&
11'2-2'343'-4'
\\
11'2-2'3-3'4-4'
&
11'2-2'34-3'-4'
&
11'2-2'34-3'4'
&
11'2-2'343'4'
\\
11'2343'2'-4'
&
11'232'-3'4-4'
&
11'2342'-3'-4'
&
11'2342'-3'4'
\\
11'2343'4'2'
&
11'2-2'3-3'44'
\end{tabular}
\end{table}
\end{ruledtabular}

\begin{ruledtabular}
\begin{table}[hbt]
\caption{
74 improper fourth order diagrams created by starting with 1-1' and
attaching one of the 32 improper third order diagrams
(first eight lines) or 42 proper third order diagrams (final eleven lines)
\label{tab.4oiL1dir}
}
\begin{tabular}{llll}
1233'2'44'-1'
&
12344'3'-1'-2'
&
1233'2'4-1'-4'
&
123-1'-2'44'-3'
\\
1233'-1'-2'44'
&
123-1'-2'-3'44'
&
122'3-1'-3'44'
&
122'34-1'-3'-4'
\\
122'344'-1'-3'
&
1233'4-1'-2'-4'
&
122'33'4-1'-4'
&
122'34-1'-3'4'
\\
122'3-1'-3'4-4'
&
1233'44'-1'-2'
&
122'343'-1'-4'
&
12344'-1'-2'-3'
\\
12344'-1'-2'3'
&
1233'-1'-2'4-4'
&
1232'44'-1'-3'
&
1232'4-1'-3'-4'
\\
12343'-1'-2'-4'
&
123-1'-2'4-3'-4'
&
123-1'-2'-3'4-4'
&
1234-1'-2'-3'-4'
\\
12343'4'-1'-2'
&
1234-1'-2'-3'4'
&
1232'3'4-1'-4'
&
1234-1'-2'3'-4'
\\
122'344'3'-1'
&
122'33'44'-1'
&
122'343'4'-1'
&
1232'3'44'-1'
\\
\hline
12344'3'2'-1'
&
12-1'-2'344'3'
&
1232'-1'-3'44'
&
12-1'-2'3-3'44'
\\
1233'42'-1'-4'
&
12-1'-2'33'4-4'
&
123-1'-2'3'44'
&
1233'4-1'-2'4'
\\
12-1'-2'33'44'
&
12344'2'-1'-3'
&
123-1'-2'44'3'
&
12-1'-2'344'-3'
\\
12343'2'-1'-4'
&
12-1'-2'343'-4'
&
1232'-1'-3'4-4'
&
12-1'-2'3-3'4-4'
\\
12342'-1'-3'-4'
&
12-1'-2'34-3'-4'
&
123-1'-2'3'4-4'
&
1234-1'-2'4'-3'
\\
12342'3'-1'-4'
&
1232'43'-1'-4'
&
12342'4'-1'-3'
&
123-1'-2'43'-4'
\\
1234-1'-2'3'4'
&
123-1'-2'43'4'
&
123-1'-2'4-3'4'
&
12343'-1'-2'4'
\\
1234-1'-2'4'3'
&
1232'4-1'-3'4'
&
12-1'-2'34-3'4'
&
12-1'-2'343'4'
\\
12344'2'3'-1'
&
1232'44'3'-1'
&
12342'-1'-3'4'
&
1233'44'2'-1'
\\
1233'42'4'-1'
&
12342'3'4'-1'
&
12342'4'3'-1'
&
12343'2'4'-1'
\\
1232'43'4'-1'
&
12343'4'2'-1'
\end{tabular}
\end{table}
\end{ruledtabular}

\subsubsection{Leading Second Order}
\label{sec.4o2l}
This set contains $N_p(2)N(2)=6\times 10$ diagrams since we have 6 proper second
order diagrams and 10 second order diagrams. These six groups
form Table \ref{tab.4oiL2}.
\begin{ruledtabular}
\begin{table}[hbt]
\caption{60 improper fourth order diagrams created by starting with
any of the six proper second order diagrams.
\label{tab.4oiL2}
}
\begin{tabular}{llll}
122'1'343'4'
&
122'1'34-3'4'
&
122'1'3-3'44'
&
122'1'3-3'4-4'
\\
122'1'344'3'
&
122'1'343'-4'
&
122'1'344'-3'
&
122'1'33'44'
\\
122'1'34-3'-4'
&
122'1'33'4-4'
\\
\hline

121'2'344'3'
&
121'2'3-3'44'
&
121'2'3-3'4-4'
&
121'2'343'4'
\\
121'2'344'-3'
&
121'2'33'44'
&
121'2'33'4-4'
&
121'2'34-3'-4'
\\
121'2'34-3'4'
&
121'2'343'-4'
\\
\hline

12-1'33'-2'44'
&
12-1'33'-2'4-4'
&
1232'-1'44'-3'
&
1233'2'-1'44'
\\
1232'3'-1'44'
&
123-1'44'-2'3'
&
1233'-1'44'-2'
&
122'33'-1'44'
\\
123-1'44'-2'-3'
&
122'3-1'44'-3'
\\
\hline

12344'3'-1'2'
&
12343'4'-1'2'
&
123-1'2'-3'44'
&
12343'-1'2'-4'
\\
123-1'2'-3'4-4'
&
1234-1'2'-3'4'
&
1233'44'-1'2'
&
12344'-1'2'-3'
\\
1233'4-1'2'-4'
&
1234-1'2'-3'-4'
\\
\hline

12-1'3-2'44'-3'
&
12-1'3-2'4-3'-4'
&
1233'2'-1'4-4'
&
122'3-1'4-3'-4'
\\
1232'-1'4-3'-4'
&
1232'3'-1'4-4'
&
123-1'4-2'3'-4'
&
122'33'-1'4-4'
\\
1233'-1'4-2'-4'
&
123-1'4-2'-3'-4'
\\
\hline

121'344'3'-2'
&
121'343'4'-2'
&
121'3-2'-3'44'
&
121'34-2'-3'4'
\\
121'3-2'-3'4-4'
&
121'343'-2'-4'
&
121'33'44'-2'
&
121'344'-2'-3'
\\
121'33'4-2'-4'
&
121'34-2'-3'-4'
\end{tabular}
\end{table}
\end{ruledtabular}

\subsubsection{Leading Third Order}
\label{sec.4o3l}
This set contains $N_p(3)N(1)=42\times 2$ diagrams since we have 42 proper third
order diagrams and 2 first order diagrams.
Those that end on the
first order exchange diagram have name patterns
\ldots 22'-\ldots,
\ldots 33'-\ldots, or
\ldots 44'-\ldots,
and are listed in Table \ref{tab.4oiL3ex}.
Those that end on the
first order tadpole diagram have name patterns \ldots 2-\ldots -2',
\ldots 3-\ldots -3' or \ldots 4'-\ldots -4
and are listed in Table \ref{tab.4oiL3dir}.

\begin{ruledtabular}
\begin{table}[hbt]
\caption{
42 improper fourth order diagrams created by starting with a proper
third order diagram and attaching the 11' diagram.
\label{tab.4oiL3ex}
}
\begin{tabular}{llll}
122'31'44'-3'
&
122'344'-1'3'
&
1233'1'44'-2'
&
121'32'44'-3'
\\
121'344'-2'3'
&
12344'-1'3'-2'
&
12344'-1'3'2'
&
1232'44'-1'3'
\\
1231'3'44'-2'
&
1231'2'44'-3'
&
12344'-1'2'3'
&
1233'2'1'44'
\\
1232'1'44'-3'
&
1231'44'-2'-3'
&
1231'44'-2'3'
&
122'33'1'44'
\\
1233'1'2'44'
&
121'33'2'44'
&
122'31'3'44'
&
1232'1'3'44'
\\
1231'3'2'44'
&
121'32'3'44'
&
1231'2'3'44'
&
1232'3'1'44'
\\
1233'-1'2'44'
&
1233'-1'44'2'
&
1233'-1'42'4'
&
1233'-1'4-2'4'
\\
1233'-1'42'-4'
&
1233'-1'2'4-4'
&
121'33'-2'4-4'
&
121'33'-2'44'
\\
122'-1'344'3'
&
122'-1'3-3'44'
&
122'-1'33'4-4'
&
122'-1'33'44'
\\
122'-1'344'-3'
&
122'-1'343'-4'
&
122'-1'3-3'4-4'
&
122'-1'34-3'-4'
\\
122'-1'34-3'4'
&
122'-1'343'4'
\end{tabular}
\end{table}
\end{ruledtabular}

\begin{ruledtabular}
\begin{table}[hbt]
\caption{
42 improper fourth order diagrams created by starting with a proper
third order diagram and attaching the 1-1' diagram.
\label{tab.4oiL3dir}
}
\begin{tabular}{llll}
121'3-2'44'-3'
&
1233'2'1'4-4'
&
12-1'344'3'-2'
&
12-1'3-2'-3'44'
\\
122'31'4-3'-4'
&
12-1'33'4-2'-4'
&
122'33'1'4-4'
&
123-1'2'44'-3'
\\
122'34-1'3'-4'
&
12-1'33'44'-2'
&
1233'1'2'4-4'
&
121'33'2'4-4'
\\
1233'1'4-2'-4'
&
122'31'3'4-4'
&
123-1'44'2'-3'
&
12-1'344'-2'-3'
\\
1232'1'4-3'-4'
&
12-1'343'-2'-4'
&
121'3-2'4-3'-4'
&
12-1'3-2'-3'4-4'
\\
1231'4-2'-3'-4'
&
12-1'34-2'-3'-4'
&
123-1'2'4-3'-4'
&
1234-1'3'-2'-4'
\\
1231'2'4-3'-4'
&
121'32'4-3'-4'
&
1231'3'4-2'-4'
&
123-1'42'-3'-4'
\\
123-1'42'4'-3'
&
1232'1'3'4-4'
&
123-1'4-2'4'-3'
&
121'34-2'3'-4'
\\
1231'3'2'4-4'
&
1234-1'3'2'-4'
&
1232'4-1'3'-4'
&
121'32'3'4-4'
\\
1234-1'2'3'-4'
&
1231'2'3'4-4'
&
12-1'34-2'-3'4'
&
12-1'343'4'-2'
\\
1231'4-2'3'-4'
&
1232'3'1'4-4'
\end{tabular}
\end{table}
\end{ruledtabular}

\subsection{proper}
Some graphs are created by replacing internal Green's functions
in $i$-th order diagrams---which become the skeleton of the
$n$-th order diagram ($i<n$)---by one or more diagrams with
a total order of $i-n$.

\subsubsection{First Order Skeleton}
\label{sec.4o1s}
74 proper fourth order diagrams are created by moving with any of the
third order diagrams into the central Green's function of the 11'
diagram, creating Table \ref{tab.4op1skelex}. The name pattern is 1\ldots 1' or 1\ldots 1'-\ldots .

\begin{ruledtabular}
\begin{table}[hbt]
\caption{
74 proper fourth order diagrams from moving a third order diagram
into 11'.
\label{tab.4op1skelex}
}
\begin{tabular}{llll}
12344'3'2'1'
&
121'-2'3-3'44'
&
121'-2'344'3'
&
1233'2'41'-4'
\\
1233'2'44'1'
&
122'344'3'1'
&
1232'1'-3'44'
&
1231'-2'44'-3'
\\
1233'1'-2'44'
&
122'31'-3'44'
&
1231'-2'-3'44'
&
1233'42'1'-4'
\\
121'-2'33'4-4'
&
1233'41'-2'-4'
&
122'33'41'-4'
&
122'344'1'-3'
\\
122'341'-3'-4'
&
1233'44'2'1'
&
1231'-2'3'44'
&
1233'41'-2'4'
\\
122'343'1'-4'
&
1232'44'1'-3'
&
121'-2'33'44'
&
12344'2'3'1'
\\
1232'44'3'1'
&
12344'2'1'-3'
&
1233'1'-2'4-4'
&
1233'44'1'-2'
\\
122'33'44'1'
&
1233'42'4'1'
&
122'31'-3'4-4'
&
1231'-2'44'3'
\\
121'-2'344'-3'
&
122'341'-3'4'
&
12344'1'-2'-3'
&
12344'1'-2'3'
\\
1232'3'44'1'
&
12343'2'1'-4'
&
121'-2'343'-4'
&
1232'41'-3'-4'
\\
12343'1'-2'-4'
&
1232'1'-3'4-4'
&
121'-2'3-3'4-4'
&
1231'-2'4-3'-4'
\\
1231'-2'-3'4-4'
&
12342'1'-3'-4'
&
121'-2'34-3'-4'
&
12341'-2'-3'-4'
\\
1231'-2'3'4-4'
&
12341'-2'4'-3'
&
12342'3'1'-4'
&
1232'43'1'-4'
\\
12342'4'1'-3'
&
1231'-2'43'-4'
&
12343'4'1'-2'
&
12341'-2'-3'4'
\\
12341'-2'3'-4'
&
1232'3'41'-4'
&
12342'1'-3'4'
&
121'-2'34-3'4'
\\
12343'4'2'1'
&
121'-2'343'4'
&
12342'3'4'1'
&
12341'-2'3'4'
\\
12341'-2'4'3'
&
12343'2'4'1'
&
12342'4'3'1'
&
1231'-2'43'4'
\\
1232'41'-3'4'
&
1232'43'4'1'
&
12343'1'-2'4'
&
1231'-2'4-3'4'
\\
12344'3'1'-2'
&
122'343'4'1'
\end{tabular}
\end{table}
\end{ruledtabular}

74 proper fourth order diagrams are created by moving with any third order
diagram into the tadpole's head of the 1-1' diagram,
tabulated in Table \ref{tab.4op1skeldir}.
The name pattern is 1-1'\ldots .

\begin{ruledtabular}
\begin{table}[hbt]
\caption{
74 proper fourth order diagrams from moving a third order diagram
into 1-1'.
\label{tab.4op1skeldir}
}
\begin{tabular}{llll}
1-1'2344'3'2'
&
1-1'2-2'344'3'
&
1-1'233'2'4-4'
&
1-1'233'2'44'
\\
1-1'22'344'3'
&
1-1'2344'3'-2'
&
1-1'232'-3'44'
&
1-1'2-2'3-3'44'
\\
1-1'23-2'44'-3'
&
1-1'233'-2'44'
&
1-1'22'3-3'44'
&
1-1'23-2'-3'44'
\\
1-1'233'42'-4'
&
1-1'2-2'33'4-4'
&
1-1'233'4-2'-4'
&
1-1'22'33'4-4'
\\
1-1'22'344'-3'
&
1-1'22'34-3'-4'
&
1-1'233'44'2'
&
1-1'23-2'3'44'
\\
1-1'233'4-2'4'
&
1-1'2-2'33'44'
&
1-1'2344'2'3'
&
1-1'232'44'3'
\\
1-1'2344'2'-3'
&
1-1'22'343'-4'
&
1-1'233'44'-2'
&
1-1'22'33'44'
\\
1-1'233'42'4'
&
1-1'23-2'44'3'
&
1-1'2-2'344'-3'
&
1-1'22'3-3'4-4'
\\
1-1'22'343'4'
&
1-1'22'34-3'4'
&
1-1'232'44'-3'
&
1-1'233'-2'4-4'
\\
1-1'2344'-2'-3'
&
1-1'2344'-2'3'
&
1-1'232'3'44'
&
1-1'2343'2'-4'
\\
1-1'2-2'343'-4'
&
1-1'232'4-3'-4'
&
1-1'2343'-2'-4'
&
1-1'232'-3'4-4'
\\
1-1'2-2'3-3'4-4'
&
1-1'23-2'4-3'-4'
&
1-1'23-2'-3'4-4'
&
1-1'2342'-3'-4'
\\
1-1'2-2'34-3'-4'
&
1-1'234-2'-3'-4'
&
1-1'23-2'3'4-4'
&
1-1'234-2'4'-3'
\\
1-1'2342'3'-4'
&
1-1'232'43'-4'
&
1-1'2342'4'-3'
&
1-1'23-2'43'-4'
\\
1-1'2343'4'-2'
&
1-1'234-2'-3'4'
&
1-1'234-2'3'-4'
&
1-1'232'3'4-4'
\\
1-1'2342'-3'4'
&
1-1'2-2'34-3'4'
&
1-1'2343'4'2'
&
1-1'2-2'343'4'
\\
1-1'2342'3'4'
&
1-1'234-2'3'4'
&
1-1'234-2'4'3'
&
1-1'2343'2'4'
\\
1-1'2342'4'3'
&
1-1'23-2'43'4'
&
1-1'232'4-3'4'
&
1-1'232'43'4'
\\
1-1'2343'-2'4'
&
1-1'23-2'4-3'4'
\end{tabular}
\end{table}
\end{ruledtabular}

\subsubsection{Second Order Skeleton}
\label{sec.4o2s}

More proper fourth order diagrams are created by introducing any of the 10 proper or improper
second order diagram into one of the three internal Green's functions of the two
proper second order diagrams of Fig.\ \ref{fig.2opnoHF}\@.
Those with skeleton 12-1'2' are in Table \ref{tab.4op2skelTP},
those with skeleton  121'2' in Table \ref{tab.4op2skelEX}\@.
Application of this method to the four second order proper
diagrams of Fig.\ \ref{fig.2opHF} generates no new diagrams beyond those
already incorporated in the previous section with first order skeletons.

\begin{ruledtabular}
\begin{table}[hbt]
\caption{30 proper fourth order diagrams by insertion of any of the 10 second order
diagrams into any of the three internal Green's functions of 12-1'2'.
\label{tab.4op2skelTP}
}
\begin{tabular}{llll}
122'33'4-1'4'
&
122'34-1'4'-3'
&
1233'4-1'4'-2'
&
1234-1'4'-2'-3'
\\
1233'2'4-1'4'
&
1232'4-1'4'-3'
&
1234-14'-2'3'
&
123-1'3'-2'4-4'
\\
123-1'3'-2'44'
&
1232'3'4-1'4'
\\
\hline
12-1'33'44'2'
&
12-1'33'42'-4'
&
12-1'344'3'2'
&
12-1'344'2'-3'
\\
12-1'342'-3'-4'
&
12-1'343'2'-4'
&
12-1'342'-3'4'
&
12-1'32'-3'4-4'
\\
12-1'32'-3'44'
&
12-1'343'4'2'
\\
\hline
12-1'2'33'44'
&
12-1'2'344'3'
&
12-1'2'3-3'44'
&
12-1'2'33'4-4'
\\
12-1'2'344'-3'
&
12-1'2'343'-4'
&
12-1'2'3-3'4-4'
&
12-1'2'34-3'-4'
\\
12-1'2'34-3'4'
&
12-1'2'343'4'
\end{tabular}
\end{table}
\end{ruledtabular}

\begin{ruledtabular}
\begin{table}[hbt]
\caption{30 proper fourth order diagrams by insertion of any of the 10 second order
diagrams into any of the three internal Green's functions of 121'2'.
\label{tab.4op2skelEX}
}
\begin{tabular}{llll}
1233'41'4'-2'
&
12341'4'-2'-3'
&
122'341'4'-3'
&
122'33'41'4'
\\
1233'2'41'4'
&
1232'41'4'-3'
&
12341'4'-2'3'
&
1231'3'-2'4-4'
\\
1231'3'-2'44'
&
1232'3'41'4'
\\
\hline
1233'44'1'2'
&
1233'41'2'-4'
&
12344'1'2'-3'
&
12341'2'-3'-4'
\\
12344'3'1'2'
&
12343'1'2'-4'
&
12341'2'-3'4'
&
1231'2'-3'4-4'
\\
1231'2'-3'44'
&
12343'4'1'2'
\\
\hline
121'33'44'2'
&
121'33'42'-4'
&
121'344'2'-3'
&
121'342'-3'-4'
\\
121'344'3'2'
&
121'343'2'-4'
&
121'342'-3'4'
&
121'32'-3'4-4'
\\
121'32'-3'44'
&
121'343'4'2'
\\
\end{tabular}
\end{table}
\end{ruledtabular}

Additional diagrams listed in Table \ref{tab.4op2skel2first}
are created by insertion of two first order diagrams
at two different places in the same two diagrams.
\begin{ruledtabular}
\begin{table}[hbt]
\caption{24 proper fourth order diagrams from insertion of two first-order diagrams at
two different places of 12-1'2' (first 12 entries) or of 121'2' (second 12 entries).
\label{tab.4op2skel2first}
}
\begin{tabular}{llll}
12-1'33'2'44'
&
12-1'32'44'-3'
&
122'3-1'44'3'
&
122'3-1'43'-4'
\\
123-1'44'3'-2'
&
123-1'43'-2'-4'
&
123-1'3'44'-2'
&
123-1'3'4-2'-4'
\\
122'3-1'3'44'
&
122'3-1'3'4-4'
&
12-1'33'2'4-4'
&
12-1'32'4-3'-4'
\\
\hline
1233'1'44'2'
&
122'31'44'3'
&
1233'1'42'-4'
&
12344'1'3'-2'
\\
1231'44'3'-2'
&
1231'44'2'-3'
&
12341'3'-2'-4'
&
1231'43'-2'-4'
\\
1231'42'-3'-4'
&
122'341'3'-4'
&
122'31'43'-4'
&
122'344'1'3'
\end{tabular}
\end{table}
\end{ruledtabular}

\subsubsection{Third Order Skeleton}

Each of the 10 diagrams of Fig.\ \ref{fig.3opcmplx} has 5 internal Green's
functions into which one can can insert one of the first order diagrams.
Insertion of 11' yields the diagrams of Table \ref{tab.4op3skelex},
insertion of 1-1' those of Table \ref{tab.4op3skeldir},
100 diagrams in total.

\begin{ruledtabular}
\begin{table}[hbt]
\caption{
50 proper fourth order diagrams from insertion of the first order exchange
diagram into the ``complex'' third order diagrams of Fig.\ \ref{fig.3opcmplx}.
\label{tab.4op3skelex}
}
\begin{tabular}{llll}
122'31'4-3'4'
&
122'34-1'4'3'
&
123-1'3'44'2'
&
122'343'1'4'
\\
1233'41'4'2'
&
1232'44'1'3'
&
123-1'3'2'44'
&
1231'3'44'2'
\\
1233'4-1'4'2'
&
1231'44'3'2'
&
12344'2'1'3'
&
12-1'33'42'4'
\\
12-1'32'44'3'
&
1233'1'4-2'4'
&
1233'1'42'4'
&
12344'2'-1'3'
\\
121'32'44'3'
&
122'31'43'4'
&
121'344'2'3'
&
1233'42'-1'4'
\\
12-1'32'3'44'
&
1232'-1'44'3'
&
121'3-2'3'44'
&
12-1'33'4-2'4'
\\
12-1'3-2'44'3'
&
12-1'344'-2'3'
&
1232'-1'3'44'
&
122'341'3'4'
\\
122'34-1'3'4'
&
123-1'44'2'3'
&
1231'44'2'3'
&
12344'1'2'3'
\\
123-1'2'3'44'
&
1231'2'44'3'
&
1233'4-1'2'4'
&
122'3-1'43'4'
\\
1232'1'44'3'
&
122'3-1'4-3'4'
&
121'33'4-2'4'
&
12-1'344'2'3'
\\
1233'42'1'4'
&
12-1'3-2'3'44'
&
121'3-2'44'3'
&
123-1'2'44'3'
\\
1233'41'2'4'
&
122'341'4'3'
&
123-1'44'3'2'
&
122'343'-1'4'
\\
121'33'42'4'
&
12344'1'3'2'
\end{tabular}
\end{table}
\end{ruledtabular}

\begin{ruledtabular}
\begin{table}[hbt]
\caption{
50 proper fourth order diagrams from insertion of the first order tadpole
diagram into the ``complex'' third order diagrams of Fig.\ \ref{fig.3opcmplx}.
\label{tab.4op3skeldir}
}
\begin{tabular}{lllll}
1234-1'4'3'-2'
&
123-1'3'42'-4'
&
12343'1'4'-2'
&
12341'4'2'-3'
\\
1232'41'3'-4'
&
123-1'3'2'4-4'
&
1231'3'42'-4'
&
1234-1'4'2'-3'
\\
1231'43'2'-4'
&
12342'1'3'-4'
&
12-1'342'4'-3'
&
12-1'32'43'-4'
\\
1231'4-2'4'-3'
&
1231'42'4'-3'
&
12342'-1'3'-4'
&
121'32'43'-4'
\\
1231'43'4'-2'
&
121'342'3'-4'
&
1231'4-2'-3'4'
&
12342'-1'4'-3'
\\
12-1'32'3'4-4'
&
12341'3'2'-4'
&
1232'-1'43'-4'
&
121'3-2'3'4-4'
\\
12-1'34-2'4'-3'
&
12-1'3-2'43'-4'
&
12-1'34-2'3'-4'
&
1232'-1'3'4-4'
\\
12341'3'4'-2'
&
1234-1'3'4'-2'
&
123-1'42'3'-4'
&
1231'42'3'-4'
\\
12341'2'3'-4'
&
123-1'2'3'4-4'
&
1231'2'43'-4'
&
1234-1'2'4'-3'
\\
123-1'43'4'-2'
&
1232'1'43'-4'
&
123-1'4-2'-3'4'
&
121'34-2'4'-3'
\\
12-1'342'3'-4'
&
12342'1'4'-3'
&
12-1'3-2'3'4-4'
&
121'3-2'43'-4'
\\
123-1'2'43'-4'
&
12341'2'4'-3'
&
12341'4'3'-2'
&
123-1'43'2'-4'
\\
12343'-1'4'-2'
&
121'342'4'-3'
\end{tabular}
\end{table}
\end{ruledtabular}

\subsubsection{complex}
\label{sec.4ocmpl}

The remaining 82 fourth order
diagrams could be generated by adding exchange
interactions to graphs in Figure \ref{fig.3opcmplx}, and some alternatively
by inserting an empty ``vacuum polarization'' bubble into one of the third order 
diagrams.
We are not trying to build a unique or well defined
inheritance scheme toward the third order parent diagrams.
Assuming the fourth order diagram is best displayed
with a minimum number of crossing lines, we classify them as if bubbles were
inserted, where possible, which yields Table \ref{tab.4opcmplTP}.

\begin{ruledtabular}
\begin{table}[hbt]
\caption{25 proper fourth order diagrams from insertion of a fermion loop
into an interaction line of one diagram of Fig.\ \ref{fig.3opcmplx}.
The third order reference
diagram is added in the column left to the line.
Each fourth order diagram is added at most once.
\label{tab.4opcmplTP}
}
\begin{tabular}{l|lll}
121'3-2'3' &
 1234-1'3'-2'4' & 
 121'3-2'4-3'4' &
\\
1232'-1'3' &
 1232'-1'4-3'4' &
\\
12-1'32'3' &
 12-1'3-2'43'4' &
 12-1'32'4-3'4' &
 12-1'343'-2'4'
\\
12-1'3-2'3' &
 12-1'3-2'4-3'4'
\\
123-1'3'2' &
 123-1'4-2'4'3' &
 123-1'3'4-2'4' &
 123-1'42'-3'4'
\\
123-1'2'3' &
 123-1'4-2'3'4' &
 123-1'43'-2'4' &
 123-1'2'4-3'4'
\\
1231'2'3' &
 12342'3'-1'4' &
 1231'43'-2'4' &
 1231'2'4-3'4'
\\
121'32'3' &
 12342'4'-1'3' &
 121'343'-2'4' &
 121'32'4-3'4'
\\
1232'1'3' &
 1232'43'-1'4' &
 12341'3'-2'4' &
 1232'1'4-3'4'
\\
1231'3'2' &
 12343'2'-1'4' &
 1231'3'4-2'4' &
 1231'42'-3'4'
\\
\end{tabular}
\end{table}
\end{ruledtabular}

The quickest way of visualizing the final set of fourth order proper
diagrams of Table \ref{tab.4opcmplX} is to erase the 4 and 4' from the
name, look up the equivalent third order diagram, and then re-insert the
photon line from 4 to 4'. This would generate 121'34-2'4'3' from 
121'3-2'3', for example. More disentangled views on same diagrams
with at least one fermion loop are obtained,
if an exchange interaction line within the loop is dropped, and a parent diagram is found
by lowering the higher vertex indices. In the same example of  121'34-2'4'3',
one could drop the 1 and 1', obtain an intermediate ill-composed name 234-2'4'3',
which is identified as 123-1'3'2' after decrementing each number by 1\@.
This is found in Fig.\ \ref{fig.3opcmplx}, and the exchange that bypasses
the vertex 2 (now re-named 1) can quickly be re-inserted.

\begin{ruledtabular}
\begin{table}[hbt]
\caption{Final set of 57 proper fourth order diagrams.
\label{tab.4opcmplX}
}
\begin{tabular}{llll}
1234-1'2'3'4'
&
12341'2'3'4'
&
1231'42'3'4'
&
12-1'34-2'3'4'
\\
12-1'34-2'4'3'
&
1231'4-2'3'4'
&
12342'-1'3'4'
&
123-1'42'3'4'
\\
12342'1'3'4'
&
121'34-2'3'4'
&
1234-1'4'2'3'
&
12341'3'4'2'
\\
12342'3'1'4'
&
12341'4'2'3'
&
1234-1'3'4'2'
&
12-1'342'3'4'
\\
121'342'3'4'
&
123-1'2'43'4'
&
12341'2'4'3'
&
12343'-1'2'4'
\\
12343'1'2'4'
&
1234-1'2'4'3'
&
1231'2'43'4'
&
1234-1'4'3'2'
\\
12343'2'1'4'
&
12341'4'3'2'
&
12-1'342'4'3'
&
1232'1'43'4'
\\
121'342'4'3'
&
12341'3'2'4'
&
1234-1'3'2'4'
&
1231'43'2'4'
\\
12342'-1'4'3'
&
1231'4-2'4'3'
&
123-1'43'4'2'
&
12342'1'4'3'
\\
12343'-1'4'2'
&
123-1'43'2'4'
&
121'34-2'4'3'
&
1232'-1'43'4'
\\
121'3-2'43'4'
&
1232'4-1'3'4'
&
1231'42'4'3'
&
1232'41'3'4'
\\
123-1'42'4'3'
&
123-1'3'42'4'
&
1232'4-1'4'3'
&
12343'1'4'2'
\\
1232'43'1'4'
&
1232'41'4'3'
&
12342'4'1'3'
&
1231'43'4'2'
\\
12-1'32'43'4'
&
12-1'343'2'4'
&
1231'3'42'4'
&
121'343'2'4'
\\
121'32'43'4'
\end{tabular}
\end{table}
\end{ruledtabular}

\section{Statistics of Hartree-Fock Terms}

\begin{figure}[hbt]
\includegraphics[scale=0.7]{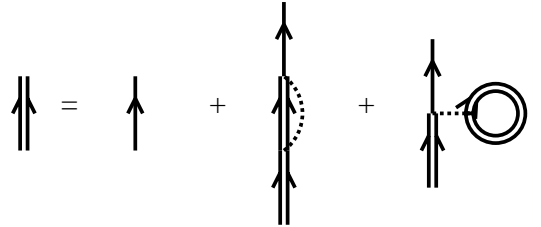}
\caption{
Hartree-Fock approximation to the Dyson equation of the Green's function
\cite[Fig.\ 10.5]{Fetter}.
\label{fig.hf}
}
\end{figure}

The Hartree-Fock (HF) approximation to the Green's function is a particular
type of merging the two first order diagrams with the Dyson equation
(Fig.\ \ref{fig.hf}).
It is equivalent
to assembling all proper graphs that can be constructed by replacing
iteratively any of the internal fermion lines in Fig.\ \ref{fig.1o} by
any of the two graphs or by any improper graphs that can
be recursively
constructed by this method. Any improper diagram that is a chain of
diagrams of this type is included as well.

In this sense, the improper second order diagrams are all of the HF-class
(Fig.\ \ref{fig.2oimp}), some of the proper diagrams are as well (Fig.\ \ref{fig.2opHF}),
and two are not (Fig.\ \ref{fig.2opnoHF}). We introduce the notation
\begin{equation}
H^{(1)}(n)=H_i^{(1)}(n)+H_p^{(1)}(n),\quad H^{(1)}(n)\le N(n)
\label{eq.hfH1}
\end{equation}
for the number of HF diagrams of order $n$, $H_i^{(1)}(n)$ of which are improper
and $H_p^{(1)}(n)$ of which are proper, bounded by the total number
of diagrams in these sub-classes:
\begin{equation}
H_i^{(1)}(n)\le N_i(n),\quad H_p^{(1)}(n)\le N_p(n).
\end{equation}

All 8 diagrams of Fig.\  \ref{fig.3oimp1},
8 diagrams of Fig. \ref{fig.3oimp1plus2},
and
8 diagrams of Fig.\ \ref{fig.3oimp2plus1}
[$H_i^{(1)}(3)=24$ out of
$N_i(3)=32$ improper third order
diagrams] are of the HF-type.
8 diagrams of Fig.\ \ref{fig.3op2in1ex},
8 diagrams of Fig.\ \ref{fig.3op2in1dir},
and none of the diagrams of Figs.\ \ref{fig.3op1in2}--\ref{fig.3opcmplx}
[total: $H_p^{(1)}(3)=16$ out of $N_p(3)=42$ proper third order diagrams] are of HF-type.
Accumulating
both statistics,
$H^{(1)}(3)=40$ out of $N(3)=74$ (54\%) of the third order diagrams
are of HF-type.

The HF-property is volatile if one looks at improper higher order composite
diagrams.
Any non-HF diagram within the partitioning destroys it.
If all pieces are of HF-type, the property is maintained:
\begin{equation}
H_i^{(1)}(n)=\sum_{l=1}^{n-l} H_p^{(1)}(l)H^{(1)}(n-l).
\label{eq.hf1}
\end{equation}
Applied to the improper
fourth order diagrams we get:
$H_p^{(1)}(1)H^{(1)}(3)=2\times 40=80$ out of the 148
diagrams of Section \ref{sec.4o1l} are of the HF type.
$H_p^{(1)}(2)H^{(1)}(2)=4\times 8=32$ of the 60 diagrams of Section \ref{sec.4o2l} are of the HF type.
With reference to Section \ref{sec.3op}, $H_p^{(1)}(3)H^{(1)}(1)=16\times 2=32$ out of the 84
diagrams of Section \ref{sec.4o3l} are in the HF class. Adding up,
$H_i^{(1)}(4)=80+32+32=144$ out of $N_i(4)=148+60+84=292$ improper fourth order diagrams
are in the HF class.

The same devaluation occurs if proper higher order diagrams are composites
of a skeleton and lower order insertions: all elements (the skeleton
and the insertions) must be of the HF class to preserve the property.
$2\times 40=80$ out of the 148 diagrams of Section \ref{sec.4o1s} are
of the HF class. None of the $84+100+82=266$ diagrams of Sections
\ref{sec.4o2s}--\ref{sec.4ocmpl} is in the HF class. So
$H_p^{(1)}(4)=80$ out of $N_p(4)=414$ proper fourth order diagrams are in the HF class. In total,
$H^{(1)}(4)=224$ out of $N(4)=706$ (32\%) of the fourth order diagrams are in the HF class.

\begin{figure}[hbt]
\includegraphics[scale=0.7]{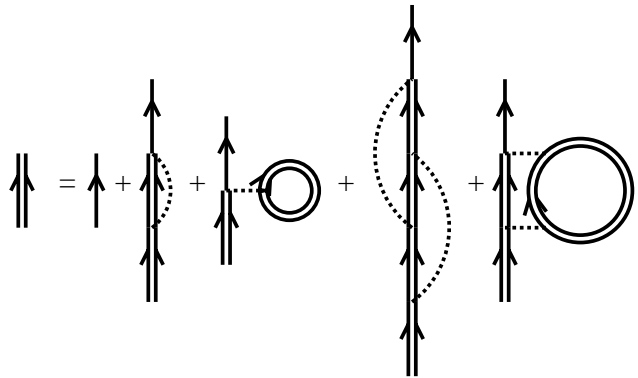}
\caption{
Extended Hartree-Fock approximation to the Dyson equation of the Green's function.
\label{fig.ehf}
}
\end{figure}

Inclusion of those second order
diagrams which are not yet part of the HF theory, those
of Fig.\ \ref{fig.2opnoHF}, extends the self-consistent field to the second
order self energy.
This
extends Fig.\ \ref{fig.hf} to Fig.\ \ref{fig.ehf},
equivalent to Cederbaum's Eq.\ (4.26) \cite{CederbaumTCA31}
and generalized in \cite[Fig.\ 4]{LindgrenPRA58}.
This extended theory includes
all second order diagrams (by construction),
all improper third order diagrams, and all proper third order
diagrams built upon a skeleton of lower order. Only the 10
third order diagrams in Fig.\ \ref{fig.3opcmplx} remain excluded.
The statistics of included third order diagrams becomes 64 out of $N(3)=74$
(86\%).
Like in (\ref{eq.hfH1}) we introduce a notation 
\begin{equation}
H^{(2)}(n)=H_i^{(2)}(n)+H_p^{(2)}(n),\, H^{(1)}(n)\le H^{(2)}(n) \le N(n)
\end{equation}
for the counts in this extended theory.
Eq.\ (\ref{eq.hf1}) and the ``devaluation rule'' remain applicable for
the upper indices raised to (2):
$H_i^{(2)}(4)=2\times 64+6\times 10+32\times 2=252$.
This evaluation yields Table \ref{tab.HFstats}.

\begin{ruledtabular}
\begin{table}[htb]
\caption{Term counts in the standard and extended HF self-consistent
field theory.
The entry 212 contains 64 diagrams of Table \ref{tab.4op1skelex},
64 diagrams of Table \ref{tab.4op1skeldir},
all of Tables \ref{tab.4op2skelTP}, \ref{tab.4op2skelEX} and
\ref{tab.4op2skel2first}, and none of Tables
\ref{tab.4op3skelex}--\ref{tab.4opcmplX}\@.
\label{tab.HFstats}}
\begin{tabular}{rrrrrr}
$n$ & $N(n)$ & $H^{(1)}(n)$ & $H^{(2)}(n)$ & $H^{(2)}_i(n)$  & $H^{(2)}_p(n)$ \\
\hline
1 & 2   &  2 & 2  & 0  & 2 \\
2 & 10  &  8 & 10 & 4  & 6 \\
3 & 74  & 40 & 64 & 32 & 32 \\
4 & 706 & 224 & 464 & 252 & 212 \\
5 & 8162 & 1344 & 3624 & 2056 & 1568 \\
6 & 110410 & 8448 & 29744 & 17336 & 12408
\end{tabular}
\end{table}
\end{ruledtabular}

\section{Higher Orders}

The diagrams of fifth and sixth order are not listed explicitly
but available in the files \texttt{public/Feynm[56][ip].txt}
on the author's web page; the first wildcard means $n$ and the second
separates proper from improper diagrams.

The series
$N(n)$
with increasing perturbative order $n$ is also obtained by insertion of
$l=0$ in \cite[(11)]{MolinariPRB71},
because this choice of a ``vertex coupling  strength'' reduces
\cite[(7)]{MolinariPRB71} to the theory of bare Coulomb
interactions. In sixth and seventh order we have
$N(6)=110410$ and $N(7)=1708394$,
see entry A000698 in the Encyclopedia of Integer Sequences \cite{EIS},
Table 2 in \cite{ArquesDMath215},
Table 1 in \cite{CvitanovicPRD18},
the list of $S_n(1)$ in \cite{TouchardCJM4},
and \cite[(2.3)]{ZinnJustinarxiv03}.
Recursive application of (\ref{eq.Ni}) yields Table \ref{tab.stats}.

\begin{ruledtabular}
\begin{table}[htb]
\caption{Term counts
and their decomposition in
subsets of improper and proper diagrams.
For comparison:
the Goldstone expansion needs 84 diagrams at $n=3$
and 3120 at $n=4$, respectively \cite{CederbaumTCA31}.
\label{tab.stats}}
\begin{tabular}{rrrr}
$n$ & $N(n)$ & $N_i(n)$ & $N_p(n)$ \\
\hline
1 & 2   &  0 & 2  \\
2 & 10  &  4 & 6 \\
3 & 74  & 32 & 42 \\
4 & 706 & 292 & 414 \\
5 & 8162 & 3104 & 5058 \\
6 & 110410 & 37924 & 72486
\end{tabular}
\end{table}
\end{ruledtabular}

\section{Summary}
The third order of the interacting Green's function expansion
in powers of the 2-fermion interaction contains 42 proper self-energy
terms, the fourth order 414 terms.

The rules for recursive construction of the names of
the $N(n)$ topologically distinct diagrams are:
(i) Start the name with a bare \texttt{1}.
(ii) Attach all possible combinations of three types of tokens, (a) bare numbers,
(b) primed numbers and (c) dashes, subject to the following constraints:
(iii) Each bare number from  1 to n appears once. This subsequence
of bare numbers is sorted in natural order from the left to the right.
(iv) Each primed number from \texttt{1'} to $n$' appears once. A primed
number appears somewhere to the right of the associated bare number.
(v) Every dash is followed by some primed number.
(vi) The set of primed numbers immediately after dashes is sorted: a
string \texttt{-}j\texttt{'} must appear after a string \texttt{-}i\texttt{'}
if $j>i$.
(vii) Within the set of primed numbers after a dash up to the next dash
or the end of the name (whichever comes first, ie, up to where
the fermion loop ends), the primed number immediately
after the dash is the smallest.

\bibliography{all}

\end{document}